\documentstyle[aas2pp4]{article}

\newcommand{\RLQ}{radio-loud quasars{}}
\newcommand{\RWQ}{radio-weak quasars{}}
\newcommand{\FIQ}{flat-spectrum radio-intermediate quasars{}}

\begin{document}
\title{The Nature of Radio-Intermediate Quasars: What is Radio-Loud
and what is Radio-Quiet?}
\author{Heino Falcke}
\affil{Astronomy Department, University of Maryland, College Park,
MD 20742-2421 (hfalcke@astro.umd.edu)}
\author{William Sherwood (p166she@mpifr-bonn.mpg.de)\and 
Alok R. Patnaik (apatnaik@mpifr-bonn.mpg.de)}
\affil{Max-Planck-Institut f\"ur Radioastronomie, Auf dem H\"ugel 69,
D-53121 Bonn, Germany}

\begin{abstract}
We have performed quasi-simultaneous radio flux density measurements
at 2.7 and 10 GHz for all PG quasars with radio flux densities between
4-200~mJy.  We find that a large fraction of these sources are
variable, flat-spectrum quasars. This brings the total fraction of
flat-spectrum quasars with a ratio between radio and optical flux of
$R>10$ --- a value previously used to define a radio-loud quasar ---
to 40\% in the PG quasar sample. We also find that the median
$R$-parameter of these flat-spectrum quasars is {\it lower} than those
of steep-spectrum radio-loud quasars. This contradicts the predictions
of the unified scheme and the idea that all flat-spectrum,
core-dominated quasars are relativistically boosted lobe-dominated
quasars. We show that this discrepancy is due to a population of
flat-spectrum radio-intermediate quasars with $25<R<250$ which can
neither be explained as relativistically boosted radio-loud quasars
nor as normal radio-weak quasars. We point out that a natural
explanation for the flat-spectrum radio-intermediate quasars is
relativistic boosting in {\it radio-weak} quasars. If the
flat-spectrum radio-intermediate quasars are considered the boosted
counterparts to usual radio-weak quasars, their fraction among
radio-weak quasars is roughly $10\%$, similar to the fraction of
boosted radio-loud quasars. This would point towards average Lorentz
factors of $\gamma_{\rm jet}=2-4$ for radio-loud and radio-weak
quasars. The presence of the flat-spectrum radio-intermediate quasars
changes the definition of 'radio-loud' and can bias some conclusions
drawn from optically selected quasar samples, where $R\simeq1-10$ is
used as the dividing line for both, flat- and steep-spectrum quasars.
Instead one should use separate $R$-parameters for the dividing line
in steep- ($R\simeq25$) and flat-spectrum ($R\sim250$) quasars.
\end{abstract}

\keywords{galaxies: active --- galaxies: jets --- galaxies: nuclei ---
galaxies: quasars --- radio continuum: galaxies}

\section{Introduction}
Since the discovery of quasars\footnote{Initially this term was used
only for radio emitting quasi-stellar objects (QSOs), but as it is
commonly done today, we will use the term 'quasar' for all QSOs} it
was known that some of them have strong radio emission and others do
not. Strittmatter et al.~(1980) showed that, in fact, there is a
dichotomy in the distribution of the radio emission of quasars, and
the studies of radio morphology of quasars have made clear why this is
so (Miller, Rawlings, \& Saunders 1993; Kellermann et al.~1994,
hereinafter K94); while radio-weak quasars show at best diffuse
extended emission, most radio-loud quasars are either point-like
or have a double-lobed radio structure, very similar to those in
Fanaroff-Riley type II (FR-II) radio galaxies. The total flux density
at the canonical radio frequency of 5~GHz of the latter is dominated
by the steep-spectrum synchrotron emission from the extended radio
lobes. Those lobes and a compact radio core in the center of \RLQ{} are
signs of a relativistic radio jet produced by the central engine of
the quasar.

This relativistic jet is also one of the keys to the unified scheme
for radio galaxies and quasars (e.g.~Barthel 1989, see also Urry \&
Padovani 1995). According to this scheme, radio galaxies,
lobe-dominated quasars, and core-dominated quasars are one and the
same type of object but seen under different aspect angles. If seen
edge-on, a dusty torus obscures the optical nucleus and the quasar is
classified as a radio galaxy. For intermediate aspect angles, the optical
nucleus, lobes and radio core are visible to give a
lobe-dominated \RLQ{}, and for face-on orientation, relativistic boosting
leads to the appearance of a core-dominated quasar, where the core is
much brighter than the lobes. A simple tool to separate core and
lobe-dominated quasars is the radio spectral index. Radio cores have a
flat and variable radio spectrum, while the lobes have a steep
spectrum and are not variable. 

The unified scheme proved to be very successful and should, in
principle, explain the radio properties of all \RLQ{}. However, it is not
clear what radio-loud and radio-weak precisely means. For this purpose
Kellermann et al.~(1989, hereinafter K89) defined the $R$-parameter as
the ratio between optical flux at 4400\AA~and radio flux density at
5~GHz and took $R=10$ as the dividing line between radio-loud and \RWQ{} (see
also K94). But as shown below this value is not necessarily the best
choice and is to some degree rather arbitrary.

For their study of the differences between radio-loud and \RWQ{}, K89 and
K94 used the PG quasar sample (Schmidt \& Green 1983), which is
optically selected and widely studied at almost all
wavelengths. Falcke, Malkan, \& Biermann (1995; hereinafter FMB95)
used the same sample and plotted the radio luminosity against the
total UV-bump luminosity (Fig.~2 in FMB95).  This diagram allows to
consider the absolute luminosity in addition to just the $R$-parameter
and highlights several things: first of all, one finds again a
separation of the two radio-loud and radio-weak populations, but there
also are a few sources which seem to be intermediate between these
two classes. FMB95 identified three sources from the low-redshift part
of the PG quasar sample which apparently were separated from the other
quasars and labeled them radio-intermediate quasars (RIQ, see also
Miller et al.~1993). What made these sources so interesting was the
fact that all three were apparently compact, variable, flat-spectrum
sources. This would have suggested that they should be
relativistically boosted \RLQ{}. However, their ratio between radio
and UV-bump luminosity was equal or even lower than that of a typical
lobe-dominated quasar -- in marked contrast to what one expects from
the unified scheme. These sources also populated the luminosity regime
$L_{\rm UV}<10^{46}$~erg/sec where no steep-spectrum \RLQ{} are found.
The suggestion FMB95 made was that these sources are, in fact, boosted
radio-weak quasars. However, it could not be excluded that they are
some rare, compact low-power radio-loud sources.

The small number of sources, which was in part due to the lack of radio
spectral information for most of the PG quasars, prevented further
conclusions.  Particularly for radio-weak and radio-intermediate
quasars only very few studies of their radio-spectra have been
published (see Barvainis, Lonsdale, \& Antonucci 1996).  We have
therefore embarked on a study to investigate the properties of the RIQ
in the PG quasar sample in more detail.

In this paper, we report quasi-simultaneous 2.7 GHz and 10.45 GHz flux
density measurements using the 100-m telescope at Effelsberg.  These
observations are used to determine the spectral indices and the
variability of the PG quasars in order to identify other flat-spectrum
RIQ. In Sec.~2, we describe the observations and present the results
in Section 3. Section 4 discusses the distribution of the
$R$-parameters of the PG quasars in light of the new results and a
summary of the paper is given in Section 5.

\section{Observations}
\subsection{Description of the observation}
The observations were performed with the MPIfR 100-m telescope in
Effelsberg on four different sessions between June 26 to June 30,
1995.  The first three nights were divided more or less equally
between 2.7 and 10.45 GHz. The fourth night was spent exclusively at
10.45 GHz.  Observations and data reduction were performed in a manner
as described in Neumann et al.~(1994).  We used the 11cm (2.695 GHz)
and 2.8cm (10.45 GHz) receivers in the secondary focus.  These
receivers can be switched within a few seconds allowing
quasi-simultaneous observations.  Other receivers were not available
during our observations. We made between 4 and 64 cross-scans for each
program source per frequency depending on the expected flux density of
the source.  On each session we observed 3C286 several times as our
primary flux calibrator.  Pointing of the telescope was frequently
monitored using nearby strong radio point sources. We had clear
weather conditions throughout the observations. Several sources were
observed at each day to look for possible intraday variability. Data
reduction was done using standard MPIfR Toolbox software (von Kap-herr
1980).

\subsection{Data Reduction}
The flux densities were scaled to the scale of Ott et al.~(1994) which
supersedes the scale by Baars et al.~(1977) and corrected with an
elevation dependent gain curve of the telescope.  The fluxes given for
each source are time averaged fluxes of all scans from the
four sessions, which were, however, all calibrated separately.  We
determined the spectral index $\alpha$ (defined as
$F_\nu\propto\nu^\alpha$) from our measured flux densities at 2.7 and
10.45 GHz.  Using the above $\alpha$ we calculated an interpolated flux
density at 4.9~GHz, which can be used to check the variability with
respect to the results of K89 --- provided the quasars have a straight
spectrum.  We estimate that our calibration error is roughly 5\% for
the brighter sources plus a statistical error (noise) of typically 2
mJy, depending on the number of scans. The final error we give in
Table 1 is the combination of statistical and calibration errors; at
2.7 GHz we are confusion limited.

\section{Results}
\subsection{The Effelsberg PG sample}
The sample we have observed was basically selected to include PG
quasars with flux densities $\ga4$ mJy at 4.9 GHz and incomplete
spectral informations. We did not observe some of the well known
\RLQ{} like e.g.~3C273 (PG
1226+023). Our final sample contained all PG quasars with $1<R<150$ (K89)
including PG 0003+19 and PG 0007+10 (III Zw 2). The results are
summarized in Table 1, where we give the radio flux densities for the
observed PG quasars. None of the sources showed intraday
variability. We therefore consider only the flux densities averaged
over the entire observing period and the spectral indices derived from
these.

The reliability of our measurements can be determined by comparing the
interpolated fluxes of the 4 steep-spectrum sources (PG 0044+03, PG
0157+00, PG 1241+17, PG 1700+51), which are not expected to be
variable, with the VLA fluxes of K89. Within our errors, the
measurements agree with each other.

\subsubsection{Spectral Indices}
Out of the 21 sources, we detected 14 at 2.7 GHz and 17 at 10 GHz.
Out of the 13 sources detected at both frequencies 6 have flat or
inverted radio spectra ($\alpha>-0.5$) between 2.7 and 10.45 GHz (PG
0007+10, PG 1309+35, PG 1333+17, PG 1538+47, PG 1718+48, PG 2209+18)
and 6 have steep spectra (PG 0044+03, PG 0157+00, PG 1241+17, PG
1351+64, PG 1407+26, PG 1700+51). One source (PG 1222+22) has
$\alpha\sim-0.5$, but due to the high flux density errors the spectral
index is highly uncertain. This means that the fraction of
flat-spectrum sources in the Effelsberg PG sample is fairly high ---
at least 30\%. Three of these flat-spectrum sources had simultaneous
spectral indices with $\alpha>0.33$ indicating synchrotron
self-absorption. These results are in agreement with the predictions
of FMB95 which implied the presence of a substantial fraction of
flat-spectrum sources with intermediate radio flux densities.

\subsubsection{Variability}
We can compare our interpolated 4.9 GHz flux densities with the K89
VLA data and find significant discrepancies for 6 sources --- these
are basically all the flat spectrum sources except PG 1333+17 but
include PG 1222+22, which may also have a flat spectrum. Those
discrepancies indicate either variability or a more complex spectral
shape --- or both.

PG 0007+10 is well known as a violently variable source
(e.g.~Ter\"asranta et al.~1992). Comparsion of our 10.45 GHz data with
those of Neumann et al.~(1994) shows that PG 1718+48 and PG
1538+47\footnote{Note that PG 1538+47 was classified as a
steep-spectrum source in FMB95, due to a non-simultaneous flux
measurement (White \& Becker 1992).} are variable sources. In
addition, comparing the K89 VLA 4.9 GHz data with literature values,
we find that also PG 1309+335 (Becker et al.~1991), PG 2209+18 (see
Machalski \& Magdziarz 1993), and PG1222-22 (Barvainis et al.~1996)
show variability. The latter authors also find clear evidence for
variability
 in PG 1216+06, PG 1407+26, and PG 1416-12.  Hence, at least
9 out of 21 sources in the Effelsberg PG sample show variability,
indicating the presence of strong compact cores at parsec scales.  The
flux of PG 1351+640 had declined in the past (Barvainis \& Antonucci
1989), but now seems to have stabilized at a lower flux-level with a
fairly steep spectrum.

\subsection{The whole PG sample}
We have supplemented the K89 radio data with our data and data from
the literature (Barvainis \& Antonucci 1989, Barvainis et al.~1996,
Neumann et al.~1994, and the NED database, see Falcke 1994). This
gives us spectral informations for all PG quasars down to an
$R$-parameter of 1 (i.e.~31 out of 113 sources). In total we have
spectral information for 49 sources. Table 2 shows the up-to-date data
list. Since the various observations were done in different epochs,
some sources can have erroneous spectral indices, due to variability.

As in FMB95, we have averaged the logarithmic flux densities at each
frequency, if multiple observations were available, and fitted a power
law to the data. Taking the geometric mean is more robust against
strong outbursts and the variability in some of the sources, moreover will we
later mainly deal with the logarithms of the fluxes. Where
appropriate, we fitted 2nd or 3rd order polynomials to the spectra in
the log-log plane. Column 2 in Table 2 gives the flux density at 4.9
GHz from those fits for all sources. For sources with only one
datapoint at 4.9 GHz we give this value.  We also tabulated the
optical flux from K89, the estimated UV-bump luminosity from FMB95 and
the differential spectral index $\alpha$ at 4.9 GHz where available. The
latter is defined as the slope of the tangential to the fitted
spectra at 4.9 GHz.

Since our time averaged flux densities differ slightly from K89 we
recalculated the $R$-parameter, which is the ratio between Col.~2 and
3 of Table 2, but also give the old K89 value (here $R_{\rm K94}$).
Moreover, we also defined a new parameter $R_{\rm UV}$, which is the
ratio between the monochromatic radio luminosity at 5~GHz in the
rest-frame and the UV-bump luminosity ($L_{\rm disk}$) divided by
$\nu=1.5\cdot10^{15}$Hz, which corresponds to a wavelength of
2000\AA{}. The wavelength was chosen such that $R_{\rm UV}$ has values
roughly similar to $R$, thus allowing an easy comparison between the
FMB95 and K89 methods of organizing the data in the optical/radio
plane.

Comparison of Col.~6 with Col.~7 \& 8 shows that all three definitions
of the $R$-parameter yield consistent results, with the exception of a
few high-redshift sources where $R_{\rm UV}$ is smaller than $R$.  In
order to make comparisons with other data sets easier, we use here the
radio/optical $R$-parameter in the observers frame for our discussion
of the radio properties of the PG quasars (Col.~7). As shown above,
taking absolute luminosities rather than fluxes would not have changed
our results significantly, that might, however, change for a
high-redshift sample.

\section{Distribution of the $R$-parameter}
\subsection{What is radio-loud?}
In the PG quasar sample, 96 out of 113 sources (85\%) have a well defined
$R$-parameter and only 17 have upper limits. Those upper limits
concern only the part of the sample where $R\la0.1$.  Since we
concentrate on the regime $R>0.1$ we will not explicitly separate the
upper limits from the detections. In Fig.~1a, we show a histogram
of the distribution of $R$-parameters for the PG quasars in logarithmic
intervals.

First of all, the bimodal distribution, which corresponds to the
radio-loud/radio-weak dichotomy is apparent and it is quite obvious
that \RWQ{} cluster around R=0.2, while \RLQ{} cluster around
R=300. Nonetheless, there are quite a few sources in between the
distributions, where radio-loud and \RWQ{} seem to blend into each
other, and it is not clear how far the tails of both distributions
reach. The classification of a quasar as radio-loud or radio-weak is,
therefore, somewhat ambiguous in the range R=10-50. If, on the other
hand, one considers only the steep-spectrum sources as shown in
Fig.~1b, the two distributions are much better separated.  The reason
for this ambiguity is the presence of a strong population of
flat-spectrum radio-intermediate quasars: six out of thirteen quasars
with $3>R>200$ have a flat radio spectrum. A similar trend was found
from the Effelsberg PG sample alone.  K94 showed that all these \FIQ{}
are compact on the VLA scale.

\subsection{Conflict with the unified scheme}
Taking $R=10$ as the dividing line between \RLQ{} and
\RWQ{} we have 22 \RLQ{} in the total sample of 113 quasars
(19\%) and 9 out of these 22 \RLQ{} are flat-spectrum sources
(41\%). This number is surprisingly high and difficult to reconcile
with the standard unified scheme for core- and lobe-dominated
sources. For a randomly oriented sample with $N$ quasars which have an
obscuring torus with semi-opening angle $\phi_{\rm opening}$ and an
average bulk Lorentz factor of $\gamma_{\rm jet}$ one finds that
\begin{equation}
n=N(1-\sqrt{1-\gamma_{\rm jet}^{-2}})/({1-\cos\phi_{\rm opening}})
\end{equation}
objects should be flat-spectrum sources assuming that objects
with inclination $i\le\arcsin\gamma_{\rm jet}^{-1}$ are core-dominated.

For $\phi_{\rm opening}=60^\circ$ and $\gamma_{\rm jet}=3$ one expects
only $10\%$ flat-spectrum sources, while a $40\%$ fraction would
indicate a Lorentz factor as low as $\gamma_{\rm jet}\simeq1.6$.  On
the other hand, for a source with $i=\gamma_{\rm jet}^{-1}$ the
Doppler enhancement of the flat-spectrum core is at best
$\gamma^3$. As the cores of steep-spectrum \RLQ{} are usually only
$10\%$ or less of the lobe luminosity at 4.9 GHz we would need at
least a 30-fold enhancement of the core flux density, hence a Lorentz
factor $\gamma_{\rm jet}\ga3$ for a radio-loud quasar to become
core-dominated.

Besides the number of sources, also the distribution of $R$-parameters
for the flat-spectrum sources is completely inconsistent with the
simple unified scheme. Obviously, core-dominated sources can only be
core-dominated as long as the core is substantially brighter than the
radio lobes. Since the lobes are expected to radiate largely
isotropically the median $R$-parameter of the flat-spectrum \RLQ{}
should always be equal or larger than the ones of the steep-spectrum
\RLQ{}. This remains true even if there is some beaming in the optical
spectrum. First of all, the optical emission would at best be boosted
by the same amount as the radio emission and secondly, the optical and
UV spectrum of the flat-spectrum PG quasars indicate that the
enhancement of the optical flux due to a possible beamed component is
still relatively modest. For example, 3C273, the source with the
strongest non-thermal beamed component, has still the largest
$R$-parameter in the sample.

In marked contrast to the expectations, the median $R$-parameter for
the steep-spectrum \RLQ{} is 217, while for the flat-spectrum \RLQ{} it
is only 94. This again is due to the large population of
\FIQ{} that cluster below the peak of the steep-spectrum
distribution. Only 2 flat-spectrum \RLQ{}, PG1226+02 (3C273) and PG
2344+09, are above the median steep-spectrum
$R$-parameter. Interestingly, only these sources occur in the right
number and have $R$-parameters expected from the unified scheme.

In summary, if one uses $R=10$ as the boundary between radio-loud and \RWQ{},
the contents of the PG quasar sample clearly contradicts the unified
scheme. There are too many flat-spectrum sources and they have on
average a radio luminosity which is too low compared with their
optical luminosity.

\subsection{Effects of a radio-selected sample}

One of the advantages of the PG quasar sample is that it is optically
selected and the radio flux densities range from several Janskys to a
fraction of milli-Jansky.  The conclusions of the above section would
be strongly altered if only a radio-selected sample is used. Typical
radio samples have a flux density cut-off of the order 1~Jy, some
deeper wide-field radio surveys go down to 200~mJy (e.g.~Patnaik et
al. 1992), however, most of the flat-spectrum \RLQ{} in the PG
quasar sample have flux densities $<200$~mJy. If we would impose a
radio flux density limit of 200~mJy for the PG quasars we would be
left with only 2 flat-spectrum and 7 steep-spectrum quasars. The
median $R$-parameters for these few sources are 1100 and 360
respectively, in reasonable agreement with the predictions of the
unified scheme. This limit is only a technical one, but illustrates how
the conclusions can be biased if the sample is radio-selected.

\subsection{Are there boosted radio-weak quasars?}
One can now ask the question, if it is possible to interpret our data
in a way that is consistent with the unified scheme rather than
contradicts it?  The clue to this answer lies in the distribution of
flat-spectrum sources themselves. As mentioned above, there does not
appear to be any problem with the unified scheme for the the high
radio flux source (i.e.~high $R$ sources), only the flat-spectrum
sources with $R\la150$ are difficult to reconcile, because of the 
presence of the \FIQ{}. This means that in order to reconcile our data
with the unified scheme, we have to give a reasonable explanation
for the nature of the \FIQ{}.

A common feature of these sources is that they have flat spectra, are
compact and are variable. This are typical characteristics of radio
cores in \RLQ{}. However, the low $R$ and the correspondingly low limit
on the extended, steep-spectrum emission is untypical for \RLQ{}.  Their
variability and their often inverted spectra argue against
Gigahertz-Peaked-Spectrum (GPS) and Compact-Steep-Spectrum (CSS)
sources, supernovae or free-free emission as an explanation.  Their
relatively large number with respect to \RLQ{} also argues against some
kind of exotic, naked \RLQ{} without radio lobes.

One viable explanation for the \FIQ{}, however, is that, just like \RLQ{},
\RWQ{} are subject to relativistic boosting and orientation effects. In
this case the \FIQ{} were just the boosted \RWQ{} and one would have two
separate bi-modal distributions for flat- and steep-spectrum
quasars, so that boosted, flat-spectrum \RWQ{} and unboosted,
steep-spectrum \RLQ{} would blend into each other, when considering only
the $R$ parameter. A proper classification of radio-loud and \RWQ{} would
then require to take $R$ {\em and} $\alpha$ into account.

In fact, classification of radio-loud and \RWQ{} can be considerably
improved in the regime $1<R<300$ if we consider steep- and
flat-spectrum sources separately. As shown in Fig.~1b a limiting value
of $R\sim25$ would effectively separate the radio-loud and the
radio-weak distributions of the steep-spectrum quasars\footnote{Here
we call the \RWQ{} with low $R$ and without spectral information also
``steep-spectrum'' quasars even though their spectral index is not
properly known. At least that we consider them to be the equivalents
to steep-spectrum radio-loud quasars in the unified scheme,
i.e.~sources which are not preferentially oriented and which make up
the bulk of the parent population.}. Only one source, PG 0044+03, a
compact steep-spectrum source with $R\simeq25$ would remain
ambiguous. All other steep-spectrum \RLQ{} do not only seem to belong
to the same distribution, but also have very similar radio
morphologies (K94), i.e.~they are all edge-brightened FR-II like radio
sources\footnote{PG 1241+17 may be a possible exception as it has one
very bright, steep-spectrum component that is connected to a second
weaker component by a bridge --- it may also be a second, peculiar CSS
quasar.}. The median $R$-parameters would then be $\left<R\right>=0.2$ for
steep-spectrum \RWQ{} and $\left<R\right>=240$ for steep-spectrum \RLQ{}.

The flat-spectrum sources are more difficult to classify as their
number is much lower. Since treating them as a single class leads to
severe inconsistencies with the unified scheme, we will make the assumption
that their distribution is bi-modal as well and shifted by
relativistic boosting to higher $R$ parameters. This implies that we
assume that \RWQ{} also have relativistic jets in their cores, which are
just a factor $\sim100-1000$ less luminous than in \RLQ{} and if pointed
towards us appear as \FIQ{}.  Unfortunately, because of the small sample,
we cannot determine the dividing line between radio-loud and
radio-weak flat-spectrum sources from Fig.~2a directly; we therefore
have to guess. 

For example, if we adopt $R=250$, which is just above the median
step-spectrum value, as the dividing $R$-parameter between the two
putative flat-spectrum distributions, the situation would change quite
significantly. Only 2 flat-spectrum sources would be considered as
radio-loud, while 10 \FIQ{} sources in the range $1<R<250$ would be
considered radio-weak. The median values for these two populations are
$\left<R\right>=1130$ and $\left<R\right>=20$ respectively. The median $R$-parameter for the
putative radio-weak flat-spectrum sources is, however, biased because
we only have complete spectral informations down to $R=1$; for
example, the median $R$-parameter for all flat-spectrum sources with
$R<250$ would be only $\left<R\right>=6.5$. It is of course possible that there
are more flat-spectrum sources at lower $R$ which have not been
observed, thus lowering $\left<R\right>$ even further. On the other hand the
limit $R=1$ is reasonable as it separates the bulk of the \RWQ{}
distribution from the radio-intermediate quasars. It may be that the
spectral index of \RWQ{} is affected by free-free emission in the nucleus
at low values of $R$. Another possible pitfall is that the extended
emission of \RWQ{} is relatively weaker than in \RLQ{} compared to their
radio cores and therefore even for larger inclinations the core still
dominates.

Anyhow, even if the assumption of a bi-modal distribution of flat-spectrum
quasars cannot yet be statistically proven, it is consistent with our
data and gives a much more satisfying interpretation of it.
The small number of {\em radio-loud} flat-spectrum ($14\%$)
sources is now in agreement with the predictions of the unified scheme
and Lorentz factors of $\gamma_{\rm jet}=2-4$. The ratio between the
median $R$-parameters for flat- and steep-spectrum \RLQ{} is $\sim5$,
which implies a relativistic boosting by a factor 50 for typical core
luminosities of $10\%$ of the total flux in steep-spectrum \RLQ{}. This is
consistent with the range of Lorentz factors mentioned above.

Moreover, with the above classification, we have a fraction of 10\%
flat-spectrum \RWQ{} with $R>1$ and a ratio of 30-100 between median
radio-weak flat- and steep-spectrum $R$-parameters. As the core
fraction in \RWQ{} is on average much higher than in \RLQ{} (i.e.~$\sim0.8$,
K89) the enhancement factors and the fraction of flat-spectrum \RWQ{} are
very similar to those of the flat-spectrum \RLQ{} and hence also
consistent with Lorentz factors of 2-4.

\section{Discussion and Summary}
We have measured quasi-simultaneous spectral indices for PG quasars
with intermediate radio luminosities and typical flux densities of
4-200 mJy. We find that a substantial fraction of these sources are
flat-spectrum sources and are variable. Six out of 13 quasars with
radio-to-optical ratios of $3>R>200$ are compact flat-spectrum
radio quasars and constitute a population of flat-spectrum
radio-intermediate quasars (RIQ).  This confirms an earlier prediction for
the radio-intermediate quasars by FMB95 which was based on a subsample
of the PG quasars.

Together with data from the literature we have now almost complete
spectral information for PG quasars down to $R\sim1$. If one uses the
definition for a radio-loud source of $R>10$ (K89) we find severe
inconsistencies in the content of the PG quasar sample with the simple
unified scheme. According to the unified scheme flat-spectrum,
core-dominated sources are the relativistically boosted counterparts
to steep-spectrum, lobe-dominated sources.  Therefore, they should be
rare and they should have higher radio flux densities than
steep-spectrum sources. However, if we treat the flat-spectrum quasars
as a single population, the flat-spectrum sources are quite frequent
($40\%$) in the PG sample and the majority has lower flux densities
and lower $R$-parameters than steep-spectrum quasars. Under these
prerequisites our observations exclude that core-dominated quasars are
generally the boosted counterparts of {\em lobe-dominated, steep-spectrum
\RLQ{}}.

We point out that this problem can be easily circumvented if one
assumes that the distributions of flat- and steep-spectrum quasars are
both bi-modal. In this case one would separate radio-loud and
radio-weak steep-spectrum sources at $R\simeq25$ and radio-loud and
radio-weak flat-spectrum quasars at a higher value of $R\sim250$,
where the latter number is fairly uncertain due to the small number of
flat-spectrum quasars.  As a consequence, the fraction of
flat-spectrum quasars would be only of the order $10\%$ for {\em
both}, radio-loud and radio-weak quasars. This fraction of
flat-spectrum sources as well as the relative enhancement of the radio
cores in radio-loud and \RWQ{} between flat- and steep-spectrum sources
is consistent with average Lorentz factors of $2-4$ and now fits well
into the unified scheme for \RLQ{}.

The implication of this suggestion is that \RWQ{} are as much subject
to relativistic boosting as are \RLQ{}, and the \FIQ{} are just the
boosted counterparts to \RWQ{} --- this agrees well with the jet-disk
symbiosis idea for radio-weak and radio-loud quasars proposed by
Falcke \& Biermann 1995 and FMB95, which postulates that the central
engines in quasars produce initially very similar radio jets. It is
known, that a large fraction of Seyfert galaxies, which are believed
to be the low-power counterparts to quasars, have collimated bi-polar
radio outflows (Ulvestad \& Wilson 1989). Radio morphological studies
of \RWQ{} also show evidence of jet-related structures (K89). Hence it
is not surprising to find jets in radio-weak quasars. However, so far
no Seyfert galaxy has shown evidence for relativistic motion in its
radio jets (although this is not yet excluded for the cores).
Therefore, the \FIQ{} could be an important clue for the understanding
of jets in radio-weak sources. It may, for example, be that the jet
speed is somehow related to the total luminosity or to the Eddington
luminosity of the central engine and becomes relativistic only for
high-power engines.

There are also a few other arguments that support the link between the
\RWQ{} and the \FIQ{}. While the \RWQ{} in the PG sample have absolute
UV luminosities that stretch from $10^{44}$erg/sec to
$10^{48}$erg/sec, steep-spectrum \RLQ{} are only found only in the
interval $10^{46}$erg/sec$<L<10^{48}$erg/sec (Falcke,
Gopal-Krishna, \& Biermann 1995). As one can infer from FMB95, the
low-redshift \FIQ{} do indeed have lower absolute powers, than the
weakest steep-spectrum \RLQ{}. With the identification of the new,
high-z \FIQ{}, the \FIQ{} now also stretch over the whole luminosity
range, just like the \RWQ{}.  Another indirect clue comes from
spectroscopic observations of $z<0.5$ PG quasars. Boroson \& Green
(1992) found that \RWQ{} have stronger \ion{Fe}{2} emission than
\RLQ{} (which they define to be at $R>1$), but they also point out
that core-dominated \RLQ{} too have relatively strong \ion{Fe}{2},
moving them closer to \RWQ{}. We note that 4 of their core-dominated 5
sources are \FIQ{} and hence might be boosted \RWQ{}.

Our interpretation can be tested further by VLBI observations of the
\FIQ{} and studies of their host galaxies. One would expect to see a
very compact nucleus in these sources, and possibly core-jet
structures and superluminal motion as seen in many lobe- and
core-dominated radio-loud quasars (e.g.~Hough et al.~1992 and Zensus,
Cohen, \& Unwin 1995). However, if the \FIQ{} are indeed radio-weak
quasars, one expects the limits for the resolved, and extended
emission (i.e.~radio lobes) to be very low, with fluxes corresponding
to $R$-parameters of unity or less. This is in marked contrast to what
one expects to see for a boosted radio-loud quasar and is a testable
prediction.  Moreover, as the host-galaxies are markedly different
between radio-loud and \RWQ{} one expects to find a substantial fraction of
spiral host galaxies for the \FIQ{}. We also need a larger optically
selected quasar sample with deep VLA observations and
quasi-simultaneous flux-measurements to directly prove the bi-modality
of flat-spectrum quasars. If these tests fail, the \FIQ{} would
constitute a major puzzle for our understanding of the radio
properties of quasars and one would have to invoke other, possibly
more exotic explanations for the \FIQ{}.

Finally we wish to point out that without spectral information the
\FIQ{} can spoil all studies concerned with the radio properties of
quasars and the difference between radio-loud and \RWQ{}. At least in
an optically selected sample one would simply overestimate the number
of radio-loud sources in certain regimes. This is especially critical
for the determination of the paucity of radio-loud quasars found at
low powers (see Falcke et al.~1995b), where some \FIQ{} could be
falsely classified as \RLQ{}.

\acknowledgements During the preparation of this manuscript we have
learned that similar observations with very similar results had been
performed by H.~Steppe and his collaborators. Due to the early death
of the PI these results have not yet been published. We thank
P.~Strittmatter for bringing this to our attention. We also thank
H.~Ter\"asranta for some more information and unpublished data on the
variability of III Zw 2, we are grateful to J\"urgen Neidh\"ofer and
Alex von Kap-herr for their friendly support and to an anonymous
referee for several suggestions and comments. This research was
supported in part by NASA under grants NAGW-3268 and NAG8-1027.

\clearpage 
\begin{deluxetable}{lrlrlrrrrr}
\small
\tablecaption{Radio flux densities and spectral indices from Effelsberg
observations}
\tablehead{
\colhead{Name}&
\multicolumn{2}{c}{2.7 GHz}&
\multicolumn{2}{c}{10.45 GHz}&
\colhead{4.9 GHz}&
\colhead{4.9 GHz}&
\colhead{$\alpha_{2.7}^{5}$}&
\colhead{$\alpha_{2.7}^{10}$}&
\colhead{$\alpha_{5}^{10}$}\\
&\multicolumn{2}{c}{mJy}&\multicolumn{2}{c}{mJy}&interp&VLA&&Eff.&}
\tablecolumns{10}
\startdata
PG 0003$+$19&$<$10& &3&$\pm$2& &3.9&$>$$-$1.6&$>$$-$0.89&$-$0.35\nl
PG 0007$+$10&102&$\pm$5.5&617&$\pm$31.&(230.)&321&1.9&1.3&0.86\nl
PG 0026$+$12&$<$10& &$<$4& & &5.1&$>$$-$1.1& &$<$$-$0.32\nl
PG 0044$+$03&78&$\pm$4.4&19&$\pm$2.6&(42.)&38&$-$1.2&$-$1.&$-$0.92\nl
PG 0157$+$00&17&$\pm$2.2&4&$\pm$2&(9.)&8&$-$1.3&$-$1.1&$-$0.92\nl
PG 0921$+$52&8&$\pm$4&$<$4& & &3.8&$-$1.2&$<$$-$0.51&$<$0.068\nl
PG 0923$+$12& & &3&$\pm$2& &10& & &$-$1.6\nl
PG 1216$+$06&$<$10& &3&$\pm$2& &4&$>$$-$1.5&$>$$-$0.89&$-$0.38\nl
PG 1222$+$22&8&$\pm$3&4&$\pm$4&(5.9)&12.&0.65&$-$0.51&$-$1.4\nl
PG 1241$+$17&306&$\pm$15.&88&$\pm$4.7&(180.)&180&$-$0.89&$-$0.92&$-$0.94\nl
PG 1309$+$35&30&$\pm$2.5&30&$\pm$2.6&(30.)&54&0.99&0&$-$0.78\nl
PG 1333$+$17&32&$\pm$2.6&21&$\pm$2.4&(27.)&25&$-$0.41&$-$0.31&$-$0.23\nl
PG 1351$+$64&24&$\pm$2.3&5&$\pm$2&(12.)&13.&$-$0.99&$-$1.2&$-$1.3\nl
PG 1407$+$26&12&$\pm$2.1&4&$\pm$2&(7.4)&7.9&$-$0.69&$-$0.81&$-$0.91\nl
PG 1416$-$12&$<$10& &$<$6& & &3.6&$>$$-$1.7& &$<$0.67\nl
PG 1538$+$47&49&$\pm$3.2&63&$\pm$3.9&(55.)&26&$-$1.1&0.19&1.2\nl
PG 1612$+$26&$<$10& &3&$\pm$1& &5.1&$>$$-$1.1&$>$$-$0.89&$-$0.69\nl
PG 1613$+$65&$<$10& &$<$4& & &3&$>$$-$2.& &$<$0.38\nl
PG 1700$+$51&15&$\pm$2.1&5&$\pm$2&(9.2)&7.2&$-$1.2&$-$0.81&$-$0.48\nl
PG 1718$+$48&112&$\pm$5.9&217&$\pm$11.&(150.)&137&0.34&0.49&0.61\nl
PG 2209$+$18&164&$\pm$164&261&$\pm$13.&(200.)&113&$-$0.62&0.34&1.1
\tablecomments{Flux densities  and spectral indices of PG quasars as measured with
the Effelsberg 100m telescope and the VLA (Kellermann et al.~1994).
All flux densities  are in mJy. Description of columns: 
(1) -- Name of source in PG sample, 
(2) -- Effelsberg 2.7 GHz flux,  
(3) -- Effelsberg 10.45 GHz flux, 
(4) -- interpolated Effelsberg 4.9 GHz flux,
(5) -- VLA 4.9 GHz flux, 
(6-8) -- spectral indices for 2.7/5, 2.7/10, and 5/10.45 GHz (only $\alpha_{2.7}^{10}$ is
simultaneous).}
\enddata
\end{deluxetable}

\clearpage
\begin{deluxetable}{lrrrrrrr}
\footnotesize
\tablecaption{Radio Properties of PG Quasars}
\tablehead{
\colhead{Name} &
\colhead{$S_\nu$(4.9 GHz)/mJy} &
\colhead{$S_{\rm o}$/mJy} &
\colhead{$\lg L_{\rm disk}$} &
\colhead{$\alpha_{\rm 5 GHz}$} &
\colhead{$R_{\rm UV}$} &
\colhead{$R$} &
\colhead{$R_{\rm K94}$}}
\startdata
0003$+$15&332.6&1.87&46.4&$-$0.8&143.&178.&175.\nl
0003$+$19&3.92&14.3&44.6&$-$0.36&0.4&0.3&0.3\nl
0007$+$10&257.6&1.63&45.4&0.66&45.2&158.&197.\nl
0026$+$12&5.1&4.74&45.5&\nodata&1.7&1.1&1.1\nl
0043$+$03&0.24&2.01&46.&\nodata&0.2&0.1&0.1\nl
0044$+$03&44.82&1.85&46.5&$-$1.&25.5&24.2&20.5\nl
0049$+$17&0.64&2.01&44.6&\nodata&0.4&0.3&0.3\nl
0050$+$12&2.43&7.94&45.&$-$0.77&0.5&0.3&0.3\nl
0052$+$25&0.74&3.08&45.6&\nodata&0.2&0.2&0.2\nl
0117$+$21&0.25&1.72&47.2&\nodata&0.1&0.1&0.2\nl
0157$+$00&7.55&3.77&45.6&$-$1.&3.&2.&2.1\nl
0804$+$76&1.29&3.94&45.4&$-$0.58&0.3&0.3&0.6\nl
0838$+$77&$<$0.15&1.37&45.&\nodata&$<$0.1&$<$0.1&0.1\nl
0844$+$34&0.31&11.4&44.9&\nodata&0.1&0.&0.\nl
0921$+$52&3.8&2.56&44.5&$-$1.2&1.&1.5&1.5\nl
0923$+$12&10.&4.83&44.3&$-$1.6&2.5&2.1&2.1\nl
0923$+$20&0.25&1.74&45.6&\nodata&0.1&0.1&0.1\nl
0934$+$01&0.53&1.38&44.3&\nodata&0.4&0.4&0.4\nl
0946$+$30&$<$0.15&1.8&47.1&\nodata&$<$0.&$<$0.1&{$<$0.1}\nl
0947$+$39&0.31&1.25&45.3&\nodata&0.3&0.2&0.3\nl
0953$+$41&1.9&4.33&46.1&\nodata&0.5&0.4&0.4\nl
1001$+$05&0.8&1.6&45.2&\nodata&0.8&0.5&0.5\nl
1004$+$13&417.9&1.92&45.8&$-$1.&225.&218.&228.\nl
1008$+$13&$<$0.15&1.45&47.7&\nodata&$<$0.&$<$0.1&{$<$0.2}\nl
1011$-$04&0.28&2.88&44.6&\nodata&0.2&0.1&0.1\nl
1012$+$00&1.17&2.&45.5&$-$0.56&0.6&0.6&0.5\nl
1022$+$51&0.37&1.61&44.1&\nodata&0.4&0.2&0.2\nl
1048$+$34&$<$0.21&2.15&45.4&\nodata&$<$0.1&$<$0.1&0.1\nl
1048$-$09&653.5&1.8&45.9&$-$0.88&494.&363.&377.\nl
1049$-$00&0.48&1.89&46.2&\nodata&0.2&0.3&0.3\nl
1100$+$77&660.&2.05&46.&$-$0.99&362.&322.&322.\nl
1103$-$00&461.9&1.77&45.9&$-$0.55&476.&261.&272.\nl
1114$+$44&0.22&1.72&45.3&\nodata&0.1&0.1&0.1\nl
1115$+$08&$<$0.15&2.09&47.6&\nodata&$<$0.&$<$0.1&{$<$0.1}\nl
1115$+$40&0.3&1.77&45.1&\nodata&0.3&0.2&0.2\nl
1116$+$21&1.97&3.87&45.9&$-$0.79&0.5&0.5&0.7\nl
1119$+$12&0.94&6.25&44.6&\nodata&0.3&0.2&0.2\nl
1121$+$42&$<$0.18&1.77&45.4&\nodata&$<$0.2&$<$0.1&0.1\nl
1126$-$04&0.51&3.05&44.9&\nodata&0.1&0.2&0.2\nl
1138$+$04&$<$0.15&1.72&47.4&\nodata&$<$0.&$<$0.1&4.8\nl
1148$+$54&1.47&2.13&47.1&$-$0.92&0.4&0.7&0.6\nl
1149$-$11&2.6&2.96&44.6&\nodata&0.9&0.9&0.9\nl
1151$+$11&$<$0.2&2.83&45.4&\nodata&$<$0.1&$<$0.1&{$<$0.1}\nl
1202$+$28&0.83&4.45&45.4&\nodata&0.5&0.2&0.2\nl
1206$+$45&$<$0.12&2.19&47.1&\nodata&$<$0.&$<$0.1&{$<$0.1}\nl
1211$+$14&0.83&6.37&45.5&\nodata&0.1&0.1&0.1\nl
1216$+$06&5.97&2.42&46.1&0.29&2.1&2.5&1.7\nl
1222$+$22&6.07&2.88&47.5&$-$0.31&1.2&2.1&4.1\nl
1226$+$02&43850.&32.5&46.5&$-$0.085&1810.&1350.&1138.\nl
1229$+$20&0.67&6.25&44.9&\nodata&0.2&0.1&0.1\nl
1241$+$17&180.9&3.19&47.2&$-$0.77&59.9&56.7&56.4\nl
1244$+$02&0.83&1.57&44.1&\nodata&0.9&0.5&0.5\nl
1247$+$26&1.&2.78&47.8&\nodata&0.1&0.4&0.4\nl
1248$+$40&$<$0.15&1.71&46.9&\nodata&$<$0.1&$<$0.1&{$<$0.1}\nl
1254$+$04&0.27&2.09&47.&\nodata&0.1&0.1&0.1\nl
1259$+$59&$<$0.15&2.16&46.3&\nodata&$<$0.1&$<$0.1&{$<$0.1}\nl
1302$-$10&913.8&4.17&45.9&0.035&385.&219.&187.\nl
1307$+$08&0.35&3.5&45.6&\nodata&0.1&0.1&0.1\nl
1309$+$35&33.76&2.99&45.5&$-$0.021&16.7&11.3&18.\nl
1310$-$10&0.26&2.73&44.4&\nodata&0.1&0.1&0.1\nl
1322$+$65&0.2&2.05&45.2&\nodata&0.2&0.1&0.1\nl
1329$+$41&0.25&1.37&47.2&\nodata&0.1&0.2&0.2\nl
1333$+$17&25.93&2.51&46.6&$-$0.31&8.2&10.3&10.\nl
1338$+$41&$<$0.15&1.67&47.3&\nodata&$<$0.&$<$0.1&{$<$0.1}\nl
1341$+$25&$<$0.23&1.92&44.7&\nodata&$<$0.2&$<$0.1&0.1\nl
1351$+$23&0.52&2.03&44.3&\nodata&0.5&0.3&0.3\nl
1351$+$64&24.15&3.08&45.2&$-$0.57&6.1&7.8&4.3\nl
1352$+$01&$<$0.15&1.75&47.2&\nodata&$<$0.&$<$0.1&{$<$0.1}\nl
1352$+$18&0.25&2.36&45.3&\nodata&0.2&0.1&0.1\nl
1354$+$21&$<$0.17&2.07&45.5&\nodata&$<$0.3&$<$0.1&0.1\nl
1402$+$26&0.62&2.68&45.3&\nodata&0.5&0.2&0.2\nl
1404$+$22&0.97&2.13&44.7&$-$0.93&1.1&0.5&0.5\nl
1407$+$26&7.04&2.31&47.&$-$0.32&1.8&3.&3.4\nl
1411$+$44&0.61&4.57&45.2&\nodata&0.2&0.1&0.1\nl
1415$+$45&0.4&2.29&44.9&\nodata&0.4&0.2&0.2\nl
1416$-$12&1.7&3.13&45.4&0.&0.6&0.5&1.1\nl
1425$+$26&132.8&2.44&46.&$-$0.75&84.4&54.4&53.6\nl
1426$+$01&1.14&4.33&45.3&$-$0.26&0.3&0.3&0.3\nl
1427$+$48&$<$0.21&1.33&45.5&\nodata&$<$0.2&$<$0.2&0.2\nl
1435$-$06&0.19&2.75&45.3&\nodata&0.1&0.1&0.1\nl
1440$+$35&1.15&4.53&44.9&$-$1.1&0.5&0.3&0.4\nl
1444$+$40&0.16&1.89&45.6&\nodata&0.1&0.1&0.1\nl
1448$+$27&1.32&4.49&44.8&$-$0.6&0.5&0.3&0.3\nl
1501$+$10&1.5&4.17&45.1&\nodata&0.1&0.4&0.4\nl
1512$+$37&351.5&1.85&46.1&$-$0.71&188.&190.&190.\nl
1519$+$22&1.5&1.66&45.&\nodata&1.5&0.9&0.9\nl
1522$+$10&0.29&2.29&47.8&\nodata&0.&0.1&0.1\nl
1534$+$58&1.92&2.75&44.5&\nodata&0.3&0.7&0.7\nl
1535$+$54&0.47&3.4&44.6&\nodata&0.1&0.1&0.1\nl
1538$+$47&49.15&1.79&46.7&$-$0.4&19.6&27.5&14.5\nl
1543$+$48&1.08&1.72&45.8&$-$0.89&1.4&0.6&0.6\nl
1545$+$21&810.9&1.72&45.9&$-$0.98&432.&471.&418.\nl
1552$+$08&0.8&1.77&44.8&\nodata&1.&0.5&0.5\nl
1612$+$26&4.87&1.8&45.7&$-$1.1&1.&2.7&2.8\nl
1613$+$65&2.5&3.03&45.4&$-$0.45&0.9&0.8&1.\nl
1617$+$17&1.89&2.63&45.2&\nodata&0.9&0.7&0.7\nl
1626$+$55&0.17&1.54&45.1&\nodata&0.1&0.1&0.1\nl
1630$+$37&$<$0.15&1.87&47.4&\nodata&$<$0.&$<$0.1&{$<$0.1}\nl
1634$+$70&1.26&4.97&47.9&$-$0.023&0.&0.3&0.4\nl
1700$+$51&7.14&3.05&46.&$-$0.95&3.6&2.3&2.4\nl
1704$+$60&1194.&1.91&45.9&$-$0.84&970.&625.&645.\nl
1715$+$53&0.8&1.37&47.4&\nodata&0.3&0.6&0.6\nl
1718$+$48&123.4&3.34&47.5&$-$0.058&8.6&36.9&41.1\nl
2112$+$05&0.91&2.81&46.5&$-$0.96&0.3&0.3&0.3\nl
2130$+$09&2.38&6.43&45.2&$-$0.51&0.4&0.4&0.3\nl
2209$+$18&188.1&2.05&44.9&0.32&62.1&91.8&141.\nl
2214$+$13&0.24&4.61&45.&\nodata&0.1&0.1&0.1\nl
2233$+$13&0.48&1.74&45.6&\nodata&0.6&0.3&0.3\nl
2251$+$11&556.8&1.43&46.&$-$0.72&324.&389.&365.\nl
2302$+$02&0.36&1.75&47.2&\nodata&0.1&0.2&0.2\nl
2304$+$04&0.77&3.02&44.4&\nodata&0.3&0.3&0.3\nl
2308$+$09&224.2&1.61&46.1&$-$0.89&184.&139.&188.\nl
2344$+$09&1523.&1.67&46.5&$-$0.18&793.&912.&1010.
\tablecomments{The columns are (1) -- source name, (2) -- average radio
flux at 5 GHz, (3) -- optical flux, (4) -- log disk luminosity in
erg/sec (FMB95), (5) -- spectral index at 5 GHz, (6) -- ratio between
radio flux and UV flux in rest frame, (7) -- ratio between radio flux
and optical flux, (8) -- same as Col.~7 but for K89 radio data}
\enddata
\end{deluxetable}

\clearpage
\figcaption[]{Distribution of the $R$-parameter, the ratio between
radio and optical flux, for all PG quasars (top) and the steep-spectrum
PG quasars (bottom). The labels directly below the bars gives the {\em
upper} cut-off for $R$ for the sources counted in that interval; the
lower cut-off is the label to the left. For most of the weak radio
sources ($R<1$) the spectrum is not known and hence they are all
marked as 'steep', even though some may have flat spectra. The
intervals are logarithmic and upper limits for R (below $R\la0.15$)
are not marked explicitly. The two distributions (radio-loud and
radio-weak) are clearly visible, but the dividing line is better
defined if one removes the flat-spectrum radio sources.}

\figcaption[]{The top panel shows the same as Fig.~1b for the
flat-spectrum PG quasars. The bottom panel shows Fig.~1a with the
flat-spectrum quasars at $R>1$ shaded in grey. Obviously a large
fraction of the sources intermediate between radio-loud and radio-weak
are flat-spectrum sources. Only a minority of two of the flat-spectrum
sources are at the upper end of the distribution as expected in the
unified scheme.}

\onecolumn
\clearpage
\begin{figure*}
\plotone{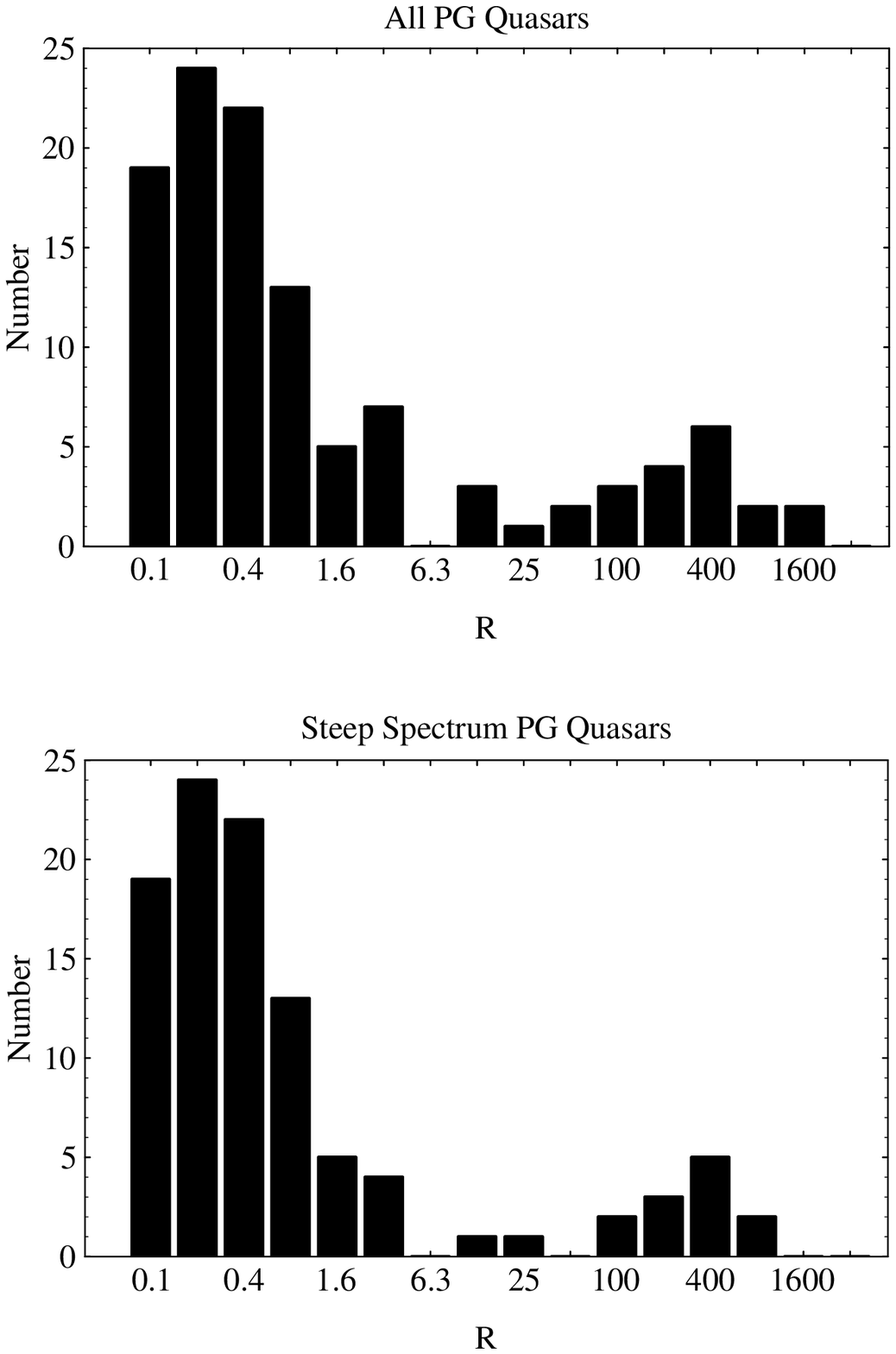}
\centerline{Figure 1}
\end{figure*}

\clearpage
\begin{figure*}
\plotone{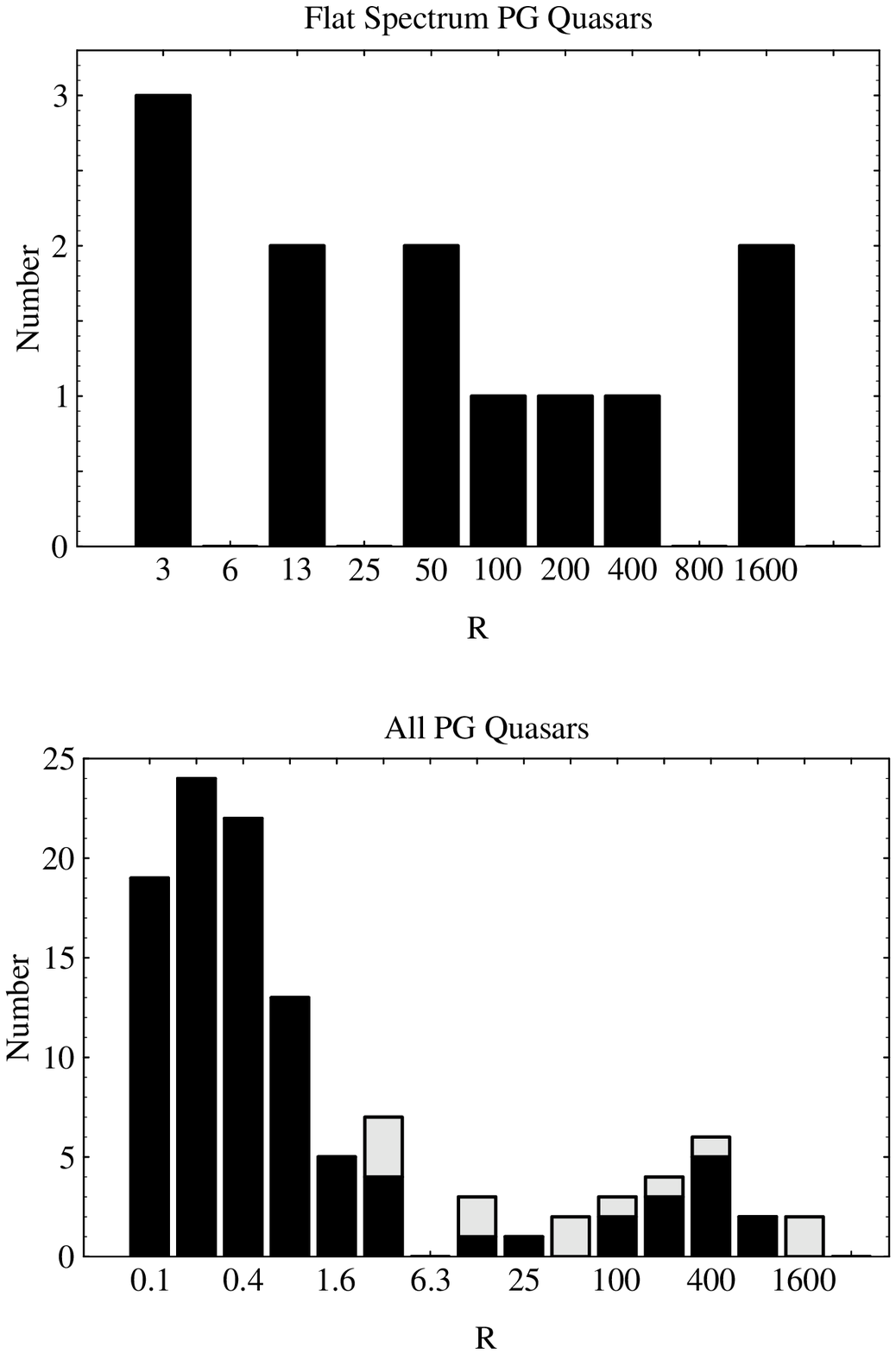}
\centerline{Figure 2}
\end{figure*}


\clearpage

\begin{thebibliography}{xxx}
\bibitem[]{}Baars, J.W.M., Genzel, R., Pauliny-Toth, I.I.K., \& Witzel,
A. 1977, 61, 99
\bibitem[]{} Barthel, P.D. 1989, ApJ 336, 606
\bibitem[]{} Barvainis, R., \& Antonucci, R. 1989, ApJS 70, 257
\bibitem[]{} Barvainis, R., Lonsdale, C., \& Antonucci, R., 1996, AJ, in press
\bibitem[]{}Becker, R.L., White, R.L., \& Edwards, A.L. 1991,
ApJS 75, 1
\bibitem[]{}Boroson, T.A., \& Green, R.F. 1992, ApJS 80, 109
\bibitem[]{}Falcke, H. 1994, PhD Thesis, RFW Universit\"at Bonn
\bibitem[]{} {Falcke, H., Mannheim, K., \& Biermann, P. L. 1993,
A\&A 278, L1} 
\bibitem[]{} {Falcke, H., \& Biermann, P.L. 1995, A\&A 293, 665}
\bibitem[]{}{Falcke, H., Malkan, M., \& Biermann, P.L. 1995a, A\&A 298,
375 (FMB95)}
\bibitem[]{}{Falcke, H., Gopal-Krishna, \& Biermann, P.L. 1995b, A\&A 298, 395}
\bibitem[]{}Hough, D.H., Readhead, A.C.S., Wood, D.A., Feldmeier, J.J.,
ApJ 393, 81 
\bibitem[]{}Kellermann, K.I., Sramek, R., Schmidt, M., Shaffer, D.B.,
\& Green, R. 1989, AJ 98, 1195 (K89)
\bibitem[]{}Kellermann, K.I., Sramek, R., Schmidt, M., Green, R., \&
Shaffer, D.B. 1995, AJ 108, 1163 (K94)
\bibitem[]{}{Miller, P., Rawlings, S., \& Saunders, R. 1993a, MNRAS 263, 425} (MRS)
\bibitem[]{}Machalski, J., \& Magdziarz, P. 1993, A\&A 267, 363
\bibitem[]{}{Neumann, M., Reich, W., F\"urst, E., Brinkmann W., Reich,
P. et al.~1994, A\&AS 106, 303}
\bibitem[]{} Ott, M., Witzel, A., Quirrenbach, A., Krichbaum, T.P.,
Standke, K.J., Schalinski, C.J., \& Hummel, C.A. 1994, A\&A 284, 331
\bibitem[]{}{Patnaik, A.R., Browne, I.W.A., Wilkinson, P.N., \& Wrobel,
J.M. 1992, MNRAS 254, 655}
\bibitem[]{}{Schmidt, M., \& Green, R. 1983, ApJ 269, 352}
\bibitem[]{}{Strittmatter, P.A., Hill, P., Pauliny-Toth, I.I.K.,
Steppe, H., \& Witzel, A. 1980, A\&A 88, L12}
\bibitem[]{}Ter\"asranta, H., Tornikoski, M., Valtaoja, E. et
al. 1992, A\&AS 94, 121
\bibitem[]{}{Ulvestad, J.S. \& Wilson, A.S. 1989, ApJ 343, 659}
\bibitem[]{}{Urry, C.M., \& Padovani, P. 1995, PASP 107, 803}
\bibitem[]{}von Kap-herr, A. 1980, Technical Report No. 40, MPIfR
Bonn, Sept. 1980
(up-dated on-line version)
\bibitem[]{}{White, R.L., \& Becker, R.H. 1992, ApJS 79, 331}
\bibitem[]{}Zensus, J.A., Cohen, M.H., Unwin, S.C. 1995, ApJ 443, 35
\end{thebibliography}
\end{document}